\documentclass[aps,prd,10pt,superscriptaddress,nofootinbib,showkeys,twocolumn]{revtex4-2}

\usepackage[T1]{fontenc}
\usepackage{stickstootext}
\usepackage[scaled=1.05]{courierten}
\usepackage[stickstoo,scaled=1.05]{newtxmath}
\usepackage[bb=dsserif,bbscaled=1.15,scr=rsfs]{mathalpha}
\usepackage{mathtools,siunitx,physics}
\usepackage{graphicx}
\usepackage[normalem]{ulem}
\usepackage{enumerate}
\usepackage{silence}
\ExplSyntaxOn
\msg_redirect_name:nnn{siunitx}{physics-pkg}{none}
\ExplSyntaxOff
\usepackage[dvipsnames]{xcolor}
\definecolor{theme}{rgb}{0.7, 0, 0.35}
\RequirePackage{hyperref}
\hypersetup{colorlinks, hypertexnames=false, allcolors=theme, bookmarksnumbered=true}%
\usepackage[capitalise]{cleveref} 
\usepackage{orcidlink}
\usepackage{xspace}

\newcommand{\etal}{\textit{et~al.}\@\xspace}
\newcommand{\ie}{i.e.,\@\xspace}

\crefname{section}{Sec.}{Secs.}
\Crefname{section}{Section}{Sections}
\crefname{appendix}{App.}{Apps.}
\Crefname{appendix}{Appendix}{Appendices}
\newcommand{\Cite}[1]{Ref.~\cite{#1}}
\newcommand{\Cites}[1]{Refs.~\cite{#1}}

\DeclareMathAlphabet{\mathpzc}{OT1}{pzc}{m}{it}

\newcommand{\mc}[1]{\mathcal{#1}}

\renewcommand{\vec}[1]{\boldsymbol{\mathrm{#1}}}

\newcommand{\sub}[1]{_\mathrm{#1}}

\graphicspath{{./}{figs/}{figures/}}

\begin{document}

\title{Are magnetic fields in cosmic voids primordial?}

\author{Deepen~Lark\,\orcidlink{0000-0001-5226-1913}}
\email{deepenlark@uni-bonn.de}
\affiliation{Argelander-Institut für Astronomie, Universität Bonn, Bonn 53121, Germany}

\author{Ruth~Durrer\,\orcidlink{0000-0001-9833-2086}}
\email{ruth.durrer@unige.ch}
\affiliation{D\'{e}partement de Physique Th\'{e}orique and Center for Astroparticle Physics, Universit\'{e} de Gen\`{e}ve,  24 quai Ernest Ansermet, 1211 Gen\`{e}ve 4, Switzerland}

\author{Jennifer~Schober\,\orcidlink{0000-0001-7888-6671}}
\email{schober@uni-bonn.de}
\affiliation{Argelander-Institut für Astronomie, Universität Bonn, Bonn 53121, Germany}

\date{\today}

\keywords{void magnetic fields, cosmological magnetic fields, primordial magnetic fields, blazars}

\begin{abstract}

The 
existence and
origin of magnetic fields in cosmic voids remains an open question:
Observational evidence inferred from gamma-ray spectra of TeV blazars has been debated for over a decade, 
while early-Universe magnetogenesis models typically invoke physics beyond the Standard Model of particle physics.
We propose a late-Universe contribution to void fields from the dipole components of galactic magnetic fields. In the conducting intergalactic medium, these fields spread through turbulent magnetic diffusion, and the galaxies residing within voids reduce the required propagation distance to a few Mpc, a scale bracketed by current estimates of the turbulent diffusivity. In the regime where the propagation distance is sufficient, the superposition of randomly oriented galactic dipoles produces volume-filling void fields with a white-noise spectrum and amplitudes well above the lower bounds inferred from the absence of GeV halos around TeV blazars. These fields constitute an astrophysical background that must be accounted for before void fields are interpreted as relics of primordial magnetogenesis, and they could account for the void fields without requiring primordial fields at all. Their white-noise spectrum provides a distinctive observational signature.

\end{abstract}

\maketitle

\section{Introduction}

The standard model of cosmology ($\Lambda$CDM) has been very successful in explaining a wide range of observations of the Universe and its history, including Hubble's law of the expansion of space \cite{book:dodelson}, numerous features of the cosmic microwave background \cite{book:durrer20}, and the large-scale structure (LSS) of the Universe. However, there are still many open questions related to, for example, the inflationary stage at the beginning \cite{ref:martin14, ref:martin16}, cosmological phase transitions that may have occurred later \cite{ref:hindmarsh21, ref:caprini20}, the asymmetry between matter and antimatter \cite{ref:canetti12}, the nature of dark matter and dark energy \cite{ref:spergel15}, the relative abundance of light elements \cite{ref:fields11}, the Hubble tension \cite{ref:divalentino21a}, the \(\sigma_8\) tension \cite{ref:divalentino21b}, galaxy formation \cite{ref:naab17}, and galactic magnetism \cite{ref:beck15,book:shukurov}.

It has been shown that magnetic fields can have an impact on several of the aforementioned phenomena and can, thus, help shape the Universe significantly (for reviews, see \Cites{ref:durrer13, ref:subramanian16, ref:vachaspati21}). Some of the leading theories of inflation \cite[see e.g.][]{ref:pajer13} predict the production of magnetic fields that can survive until today in the voids of LSS \cite{ref:anber10, my:inf_noback}. These primordial fields would also impact the spectrum of 
relic gravitational waves \cite{ref:roperpol20} from inflation \cite{ref:bamba15} as well as from (first-order) cosmological phase transitions \cite{ref:caprini09}. If primordial magnetic fields are helical as predicted by some inflationary magnetogenesis models, they can even help explain the asymmetry between matter (baryons) and antimatter (antibaryons) through the conservation of the Chern-Simons number \cite{ref:fujita16, ref:vachaspati01, tex:boyer25}. If present, primordial magnetic fields have also been shown to play a role in many important tensions in modern cosmology, like the Hubble tension and the \(\sigma_8\) tension \cite{Jedamzik:2025cax, tex:kids2025}, and the lithium abundance problem \cite{ref:yamazaki14}.
Beyond their direct effect on the physics of the early Universe, primordial magnetic fields can influence how cosmic structures form in the post-recombination Universe. 
Through additional density fluctuations in the primordial plasma, they modify the seeds of cosmic structure formation, which affects galaxy formation \cite{ref:sanati24} and the reionization epoch in the late Universe, and therefore the global 21cm signal \cite{ref:Schleicher09,ref:Sethi09,ref:Kunze19}.

Beyond their significance in astrophysics, primordial magnetic fields also have implications for fundamental physics at microscopic scales. One of our leading ways to probe fundamental high-energy physics is via particle accelerators, which currently can reach energies up to $\sim 10$ TeV scales. However, the Standard Model
of particle physics 
(SM) may well remain valid for many orders of magnitude above the current particle accelerators, which will not be accessible by accelerators in the foreseeable future \cite{ref:tanabashi18}. Consequently, exploring the physics beyond the SM (BSM) remains an uncertain task with accelerator physics. On the other hand, the early Universe can help us get insight into some of these questions, since the energy was much higher than accelerator energies (maybe even up to the Planck mass \(\simeq 10^{16}\) TeV). Since most magnetogenesis models proposed in the literature employ BSM physics, a detection of these primordial fields would open a direct window into extensions of the SM.

The high significance of primordial magnetic fields for addressing critical questions studied across different communities prompts several ongoing efforts to detect them (for a review, see \Cite{ref:batista21}). However, the only place in the present-day Universe where these primordial fields can be expected to be in their original form is in the voids of the LSS. 
In the other regions of the Universe, the signature of the primordial magnetic field is most likely
strongly
contaminated by astrophysical processes in the late Universe.
One of the main such processes is the dynamo mechanism, which amplifies magnetic fields exponentially by randomly stretching, 
twisting, and
folding magnetic field lines through turbulent plasma motion \cite{ref:brandenburg05,ref:Schober13,ref:Martin-Alvarez22}. These highly non-linear plasma processes make it practically impossible to deduce the initial conditions of the magnetic fields before the emergence of cosmic 
structures \Cite{Schober:2026cyf}, 
leaving the void fields as the only potential relic of the primordial field.

While theorized for decades, the first evidence for the existence of void fields was reported in 2010 \cite{ref:neronov10a, ref:neronov10b}, providing a proof of concept of the viability of primordial fields as a probe of fundamental physics. There, the authors used the spectra of TeV blazars to deduce the lower limit of the magnetic field strength in voids of \(10^{-16}\) G for coherence length scales above 1 Mpc. While the high volume-filling factor suggested by these observations favors a primordial origin, astrophysical processes in the late Universe like plasma instabilities \cite{ref:broderick12} and galactic winds \cite{ref:bertone06} have also been proposed to explain blazar observations and void magnetic fields, respectively. 
Given the extreme range of density and energy scales involved, both analytical and numerical studies have not yet provided a conclusive resolution, and the existence and the nature of void fields remains an open problem. A clear proof of the existence of void fields would be the observation of GeV halos around TeV blazars, which would allow for a direct measurement of the void field strength, and simultaneously rule out alternative interpretations of the missing GeV observations \cite{BroderickEtAl2018}.
Recently, an extended GeV halo around the blazar Mkn~501 has been reported, consistent with an intergalactic field of about \SI{1.5e-15}{G} and a coherence length of about 10~kpc~\cite{tex:webar25}. This combination is difficult to reconcile with a primordial origin: causally generated fields that have undergone turbulent decay end on the largest-processed-eddy relation~\cite{ref:banerjee04,ref:hosking23}, orders of magnitude above this strength at kpc scales. This points to a contribution from late-Universe processes. Whether such processes pollute primordial fields or act as the dominant magnetogenesis source, further analysis is required to elucidate the nature of cosmological magnetic fields.

The propagation of galactic magnetic fields into voids has recently been examined in several works~\cite{Ghosh:2025bqp,Brandenburg:2025utr,Seller:2025ogo,Yamashita:2026dsn}. In this article, we propose a late-Universe contribution to void fields from the dipole components of galactic magnetic fields, which could contaminate primordial fields in voids with fields of comparable or larger strength and, if the dipole fields spread over a few Mpc, could account for the void fields without primordial magnetogenesis. Specifically, we show that in this regime the superposition of galactic dipoles produces volume-filling void fields with strengths within the range suggested by the blazar spectra.

\section{Set-up and assumptions}
\label{sec:setup}

The large-scale magnetic fields of spiral galaxies are maintained by mean-field dynamos operating in the disc and, possibly, in the halo~\cite{ref:beck15,book:shukurov}.
Disc dynamos favor fields of even (quadrupolar) symmetry, whereas halo dynamos can sustain odd (dipolar) modes, so that mixed-parity configurations, quadrupolar near the disc and dipolar in the halo, arise naturally~\cite{ref:brandenburg92}.
Such configurations emerge in magnetohydrodynamic simulations of Milky-Way-like galaxies~\cite{ref:ntormousi20} and are consistent with models of the Galactic field that include a substantial out-of-plane halo component~\cite{ref:jansson12b,ref:unger24}.
In this work, we focus on the dipole component, whose field in vacuum decays as \(r^{-3}\), the slowest falloff among the multipoles; its behavior in the conducting intergalactic medium is discussed below. Because the dynamo continuously regenerates the galactic field, the dipole acts as a sustained source rather than a fixed initial field, a distinction that becomes important for its propagation. We parametrize its strength by the fraction \(f\sub{dp}\) of the galactic magnetic energy it carries, adopting \(f\sub{dp} \approx 0.01\) as a fiducial value, and show in \cref{sec:numerics} that even considerably smaller fractions suffice.
Since contributions from individual galaxies add in quadrature, \(f\sub{dp}\) should be understood as a population average: if only a fraction \(p\) of galaxies host dipolar halo fields, the results correspond to an effective \(f\sub{dp,eff} = p\, f\sub{dp}\); for example, setup $B$ in \cref{tab:input} can be understood to cover the case in which one percent of galaxies carry one percent of their magnetic energy in the dipole.

The intergalactic medium in voids is a fully ionized plasma with free electron densities of order \num{e-2}--\SI{e-1}{m^{-3}}~\cite{ref:miniati13,Ghosh:2025bqp}, and its high conductivity governs how galactic fields reach into it.
A galactic field spreads into the surrounding plasma through the motion of the plasma itself, since magnetic flux is effectively frozen into the gas on cosmological scales.
Two channels contribute: coherent galactic outflows, which carry the field along with the gas~\cite{ref:bertone06}, and the turbulent motions that pervade the cosmic web~\cite{Ryu:2008hi,ref:ayromlou24}, which spread the field relative to the mean flow and 
result in an effective turbulent magnetic diffusivity~\cite{ref:brandenburg05}.\footnote{The microscopic Spitzer resistivity~\cite{Spitzer1962} is many orders of magnitude smaller than the turbulent diffusivity in the intergalactic medium and is negligible on the scales considered here~\cite{ref:brandenburg05,Ghosh:2025bqp}.} The efficiency of magnetized outflows depends sensitively on the adopted feedback model~\cite{Garcia:2020kxm,Blunier:2024aqx}, and we therefore focus on turbulent diffusion in this work. Mean-field simulations of a galactic dynamo embedded in a turbulent exterior show that, within the region the field reaches, the dipole retains its \(r^{-3}\) falloff~\cite{Ghosh:2025bqp,Brandenburg:2025utr}. The dipole profile adopted in the following sections therefore remains valid within this region, and the question becomes how far turbulent diffusion can carry the field into voids.

Although underdense, voids host a substantial population of galaxies, both isolated and in bound groups~\cite{ref:fernandez26}. The luminosity function of void galaxies~\cite{ref:moorman15} implies a mean separation of about 5--7~Mpc once dwarf galaxies are included,
comparable to the intergalactic distance \(d\sub{IG}\) adopted below.
This estimate follows from integrating the Schechter fit down to an absolute $r$-band magnitude of about $-14$, within the measured range. Because the faint-end slope is shallow, the mean separation depends only weakly on the luminosity cutoff. Dwarf galaxies host total fields of about \SI{4e-6}{G}, generated predominantly by a small-scale dynamo~\cite{Chyzy:2011sw}; their weaker ordered components lower the population average \(f\sub{dp}\) introduced above.

For the dipole fields of these galaxies to fill a void, they must therefore spread over distances of a few Mpc, of the order of half this separation. Current estimates of the propagation distance bracket this requirement: \Cite{Ghosh:2025bqp} finds that dynamo-generated fields spread to galactocentric radii of a few hundred kpc, whereas \Cite{Yamashita:2026dsn} obtains a magnetic screening length of 6--7~Mpc. In this work, we consider the regime in which the propagation distance is sufficient for the dipole fields to overlap throughout the void, and determine the resulting void fields; a full study 
of the required effective diffusivity to sufficiently fill void with galactic fields is deferred to future work.

The void field is thus set by two factors: the distance over which turbulent diffusion carries the galactic fields, and the \(r^{-3}\) falloff of the dipole within this distance.
In this work, we set aside  the first point and ask whether the \(r^{-3}\) falloff by itself, without the plasma processes that limit the reach of the field, produces void fields of the strength suggested by the blazar spectra.
Only in this case, additional complications like the finite range of the fields, the galaxies inside the void or a limited filling fraction are worth considering.

In the following we adopt  a deliberately stringent configuration, in which galaxies are distributed over the boundary of an otherwise empty void of radius about $\SI{20}{Mpc}$ and their dipole fields extend across it. The radius of \(r\sub{void} \simeq \SI{20}{Mpc}\) is typical of large voids identified in galaxy surveys~\cite{ref:pan12} and cosmological simulations~\cite{Curtis:2025cwu}.
In this configuration, the field at the center has to bridge the full $\SI{20}{Mpc}$, several times the few Mpc required once the void galaxies are included; since these galaxies lie closer to any point in the void than its boundary does, the configuration yields a conservative estimate for given galactic parameters.
The superposed dipole fields then fill the void volume and form a background that must be accounted for before void fields are interpreted as relics of primordial magnetogenesis, much as astrophysical foregrounds must be removed from the cosmic microwave background. In the following, we first estimate the root mean square (RMS) field at the center of a void, where it is weakest, and then compute its spectrum numerically in \cref{sec:numerics}.

\section{Analytical estimate at the center of a void}

We first estimate the field at the center of a void, where it is weakest. For this order-of-magnitude estimate, we consider a spherical void of radius \(r\sub{void} \simeq \SI{20}{Mpc}\), as introduced in \cref{sec:setup}. Although voids are bounded by filaments and nodes, we adopt the idealized configuration described in \cref{sec:setup}, in which galaxies are distributed uniformly over the surface of the void; since galaxies within the void would contribute additional, randomly oriented fields, this configuration yields a conservative estimate.

For a typical density of $L^*$ galaxies comparable to the Milky Way, $0.01/$Mpc$^3$, corresponding to an intergalactic distance \(d\sub{IG} \simeq 5\)~Mpc, the number of galaxies surrounding the void is
\begin{gather}
    N\sub{g} = \frac{4\pi r\sub{void}^2}{d\sub{IG}^2} \,.
\end{gather}
We take all galaxies to have magnetic fields of similar strength, \(B\sub{g} \simeq \SI{3e-6}{G}\), with a fraction \(f\sub{dp} \simeq 0.01\) of the magnetic energy in the dipole component, so that \(B\sub{dp} \simeq \sqrt{f\sub{dp}}\, B\sub{g}\); in \cref{sec:numerics}, we draw the galactic field strengths from a lognormal distribution instead.
The dipole of each galaxy decays as \(r^{-3}\) to contribute \(B\sub{void}^{(1)}\) from one galaxy at the center of the void, given by
\begin{gather}
    B\sub{void}^{(1)} = B\sub{dp} \times \left(\frac{r\sub{dp}}{r\sub{void}}\right)^3 \,.
\end{gather}
Here, \(r\sub{dp}\) is the distance from the center of the galaxy at which \(B\sub{dp} \simeq \sqrt{f\sub{dp}} \times \SI{3e-6}{G}\).
According to \Cites{ref:unger24,ref:jansson12a,ref:korochkin25}, the out-of-plane component of the galactic magnetic field remains of the order of $\mu$G up to the distance of \(7\)~kpc, so we set \(r\sub{dp} \simeq 7\)~kpc.
This choice is conservative: in idealized mean-field simulations of a saturated dipolar dynamo, the external field follows \(B \approx 0.7\,B\sub{eq}(R/r)^3\), where \(R\) is the radius of the dynamo region and \(B\sub{eq}\) the equipartition field strength~\cite{Brandenburg:2025utr}, so that for a dynamo region of a few kpc our fiducial normalization lies a factor of a few below this value. Any non-dipolar contribution to the out-of-plane field at this radius is effectively absorbed into \(f\sub{dp}\).
We then have \(N\sub{g}\) galaxy dipoles each contributing a magnetic field of the strength \(B\sub{void}^{(1)}\) at the center of the void. In this simple estimate, we neglect possible correlations of the galactic dipole fields and assume a random orientation that is uniformly distributed across the solid angle.
Each component of \(\vec{B}\sub{void}^{(1)}\) then has zero mean and variance \((B\sub{void}^{(1)})^2/3\); for a fixed magnitude, it is uniformly distributed between \(-B\sub{void}^{(1)}\) and \(B\sub{void}^{(1)}\). Upon adding \(N\sub{g}\) such independent contributions, each component of the resulting field becomes normally distributed by the central limit theorem.
The magnitude of this total magnetic field at the center will have the Maxwell-Boltzmann distribution with the RMS given by \(B\sub{total}^{\rm rms} = B^{(1)}\sub{void} \times \sqrt{N\sub{g}}\). Combining all of this, we obtain a rough estimate for the resulting field
\begin{align}
    B\sub{total}^{\rm rms} =&\, B^{(1)}\sub{void} \times \sqrt{N\sub{g}}
    \nonumber \\
    =&\, \sqrt{f\sub{dp}} \times B\sub{g} \times \left(\frac{r\sub{dp}}{r\sub{void}}\right)^3 \times \frac{2 \sqrt{\pi} r\sub{void}}{d\sub{IG}}
    \nonumber \\
    \simeq&\, 10^{-16} \unit{G}
    \times \left(\frac{f\sub{dp}}{0.01}\right)^{1/2} \times \left( \frac{B\sub{g}}{3\times 10^{-6}~\unit{G}} \right)
    \nonumber \\
    &\, \times \left( \frac{5~\unit{Mpc}}{d\sub{IG}}\right) \times \left( \frac{r\sub{dp}}{7~\unit{kpc}} \right)^3
    \times \left( \frac{20~\unit{Mpc}}{r\sub{void}}\right)^2. 
    \label{eq_Bcenter}
\end{align}
This is the RMS field at the center of the void, where it is weakest. The field is larger elsewhere in the void, but its dependence on the galactic parameters still follows \cref{eq_Bcenter}. In the next section, we calculate the RMS value of the magnetic field numerically in the entire void and estimate its power spectrum.

\section{Spectrum of the magnetic field in the whole void}
\label{sec:numerics}

In order to calculate the RMS magnetic field from these dipoles and its coherence scale in the whole void numerically, we start with the relatively simple configuration assumed in the previous section, \ie of a totally empty void with galaxies uniformly distributed over its surface. For simplicity, we consider a cubic void rather than a spherical one in the numerical analysis, with the assumption that the void is sufficiently large that its shape does not impact the magnetic fields appreciably in most of the void volume. The void is modeled as a $512^3$ cube with galaxies spread on the surface of the cube as point particles. The number of galaxies is taken to be 204, which is approximately the number of galaxies expected around a spherical void with \(r\sub{void} \simeq\) 20 Mpc and 
\(d\sub{IG} \simeq 5~\unit{Mpc}\). 
The cube has the same surface area as the spherical void, giving a side length \(d\sub{IG}\sqrt{N\sub{g}/6} \simeq 29\)~Mpc and a cell size of about 57~kpc on the \(512^3\) grid. The field at each grid point is computed as the direct sum of the point-dipole fields, without periodic images, and the power spectrum is obtained from a fast Fourier transform of this field without padding or windowing; the implied periodicity mainly affects the largest scales, while the spectrum on smaller scales is dominated by the point-like sources.

Each galaxy is assumed to have a magnetic field of random strength with a lognormal probability distribution with median \(\mu\sub{B}\)
\begin{gather}
    \ln \left(\frac{B\sub{g}}{\mu\sub{B}}\right) \sim \mc{N} \left(0, \sigma^2\right) \,.
    \label{eq:bg}
\end{gather}
As discussed in \cref{sec:sigmaB}, the expected strength of the void field depends on the width of this distribution, scaling as \(\mu\sub{B}\,e^{\sigma^2}\). Each galactic field is assumed to have a dipole component of strength \(B\sub{dp} = \sqrt{f\sub{dp}}\,B\sub{g}\) at a distance \(r\sub{dp}\) from the center of the galaxy, with an orientation drawn uniformly over the sphere by normalizing vectors of independent Gaussian components~\cite{Muller1959}. We consider three sets of values for these parameters, listed in \cref{tab:input}.
\begin{table}
\begin{ruledtabular}
\begin{tabular}{lccc}
Parameter                        & Setup A & Setup B & Setup C \\
\colrule \noalign{\vspace{4pt}}
$f\sub{dp}$                      & $10^{-2}$   & $10^{-4}$  & $10^{-1}$  \\ \noalign{\vspace{4pt}}
$r\sub{dp} (\unit{kpc})$         & 7 & 2 & 10 \\ \noalign{\vspace{4pt}}
$\sigma$                         & 1 & 0 & 2 \\ \noalign{\vspace{4pt}}
$\mu\sub{B} (\unit{\mu G})$      & 3 & 1 & 10
\end{tabular}
\end{ruledtabular}
\caption{Parameter values for the three setups shown in \cref{fig:Bfield}, with \(d\sub{IG} = 5\unit{Mpc}\) and \(r\sub{void} = 20\unit{Mpc}\) fixed throughout. Setup $A$ is our fiducial case. Setups $B$ and $C$ combine lower and higher values of all parameters, respectively, to illustrate the width of the range of possible void field strengths; the dependence on each individual parameter follows from \cref{eq_Bcenter}, and the dependence on \(\sigma\) from \cref{sec:sigmaB}.}
\label{tab:input}
\end{table}

The power spectrum \(P_B(k)\) of the resulting fields, defined as
\begin{gather}
    (2\pi)^3 \delta^{(3)} (\vec k - \vec k') P\sub B (k) = \frac{1}{2}\left \langle \vec{B}(\vec k) \cdot\vec{B}^\ast (\vec k') \right\rangle \,,
    \\
    \intertext{where}
    \vec B (\vec k) = \int \dd^3x \, \vec B (\vec x) e^{i \vec k \cdot \vec x} \,,
\end{gather}
is calculated from the average of the spectrum over multiple ensembles. Details about the ensemble averaging and the convergence of the results are reported in \cref{sec:convergence}.

\begin{figure}
    \centering
    \includegraphics[width=\linewidth]{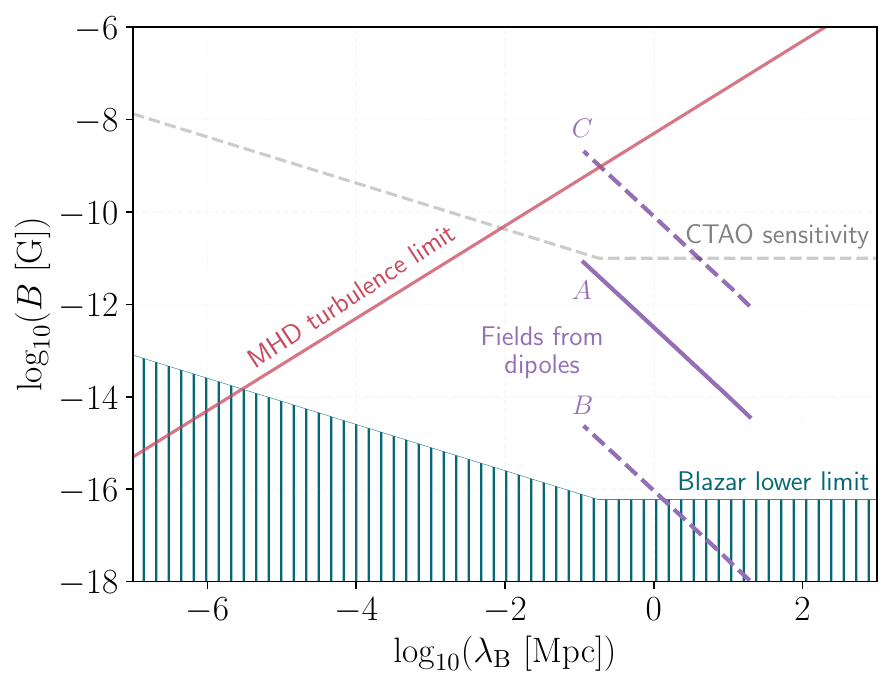}
    \caption{The characteristic magnetic field (purple lines), as defined in \cref{ePspec1}, from the galactic dipoles on different length scales ($\lambda_B=2\pi/k$) for a cube void of size \(\simeq\) 29 Mpc surrounded by 204 dipoles uniformly distributed over the surface of the void; three different setups $A$, $B$, and $C$ detailed in \cref{tab:input} are shown. More details can be found in the paragraph after \cref{ePspec1}.
    }
    \label{fig:Bfield}
\end{figure}

The resulting spectrum is that of white noise, \(P\sub{B}(k) \sim k^0\), as expected for uncorrelated point-like sources.
The Fourier transform of a dipole field is independent of \(k\), so on scales much larger than the grid spacing the spectrum is set by the dipole moments alone and is insensitive to the numerical resolution. We have verified that simulations with \(256^3\) and \(512^3\) grid points yield consistent spectra.
The characteristic magnetic field for a given coherence length scale \(\lambda_B = 2\pi/k\) is then defined as 
\begin{equation}
    \label{ePspec1}
    B\left(\lambda\sub{B}\right) = \frac{\sqrt{P\sub{B} (k) k^3}}{\pi} \,.
\end{equation}
Throughout, the galactic input parameters (\(B\sub{g}\), \(B\sub{dp}\), and \(r\sub{dp}\)) denote typical values, while the resulting void fields are RMS quantities.
This characteristic field across various length scales is shown in \cref{fig:Bfield} as purple lines, with the solid line denoting the results for setup $A$ and the dashed lines denoting setups $B$ and $C$. For setup $A$, the characteristic field at the scale of 1~Mpc is around \(\SI{3e-13}{G}\), about three and a half orders of magnitude stronger than the rough estimate of the field at the center of the void \eqref{eq_Bcenter}.
This difference is expected, since the volume-averaged field is dominated by regions closer to the galaxies, whereas \cref{eq_Bcenter} applies at the center, where the field is weakest.
We show the spectrum down to about \SI{100}{kpc}, below which our model, which treats galaxies as point dipoles, cannot capture the relevant physics.
The lower hatched region (blue with vertical hatching) in \cref{fig:Bfield} represents the region excluded by the earlier observations of TeV blazars, which provides a lower limit for void fields \cite{ref:neronov10b}. It has been suggested in \Cite{ref:caprini15} that, for smaller scales, the lower limit could have a different slope or shift to the right depending on the spectral index of the magnetic fields present. The diagonal red line denotes the size of the largest processed eddy at recombination, and, thus, represents the approximate region where primordial fields could be expected \cite{ref:banerjee04, ref:hosking23}. Finally, the grey dashed line represents the theorized sensitivity below which Cherenkov Telescope Array Observatory (CTAO) is expected to be able to detect the void magnetic fields \cite{tex:boyer25}.

As shown in \cref{fig:Bfield}, the slope of the spectrum of the magnetic fields remains that of a white-noise spectrum for the different setups given in \cref{tab:input}.
Correlations among the galactic dipoles, for example from intrinsic alignments of galaxy spins, could modify the spectral shape; such effects are beyond the scope of this article and are deferred to future work.
The absolute value of the field strength, on the other hand, changes between the setups. Setups $B$ and $C$ illustrate the width of the range of possible field strengths. The dependence on the galactic parameters is the same as given by \cref{eq_Bcenter} with two exceptions. First, the amplitude would change from 
\(10^{-16}~\unit{G}\) 
to the scale-dependent value depicted by the solid purple line in \cref{fig:Bfield}. Second, the linear dependence on the galactic field \(B\sub{g}\) in \cref{eq_Bcenter} is replaced by a dependence on \(\mu\sub{B}\,e^{\sigma^2}\), the RMS galactic field for the lognormal distribution of \cref{eq:bg}, as derived in \cref{sec:sigmaB}.

The extended halo reported around Mkn~501~\cite{tex:webar25} corresponds to a field of about \SI{1.5e-15}{G} with a coherence length of about 10~kpc. Its strength lies within the range spanned by our setups on the scales shown in \cref{fig:Bfield}, but a quantitative comparison is not straightforward, since this coherence length lies below the scales described by our model and the inferred field strength depends on the spectral shape assumed in the cascade modeling. Measurements of the spectral slope of void fields would therefore provide a direct test of the dipole contribution.

\section{Conclusion}

Our estimates show that, if galactic dipole fields spread over distances comparable to the separation of void galaxies, their superposition produces void fields well above the lower bounds inferred from the absence of GeV halos around TeV blazars.

The assumption of a dipole component of the galactic magnetic field is motivated by models of the Milky Way field~\cite{ref:jansson12b,ref:unger24} and by simulations of galactic halos~\cite{ref:ntormousi20}, which show that an \(\alpha\Omega\) dynamo can produce dipole fields among other components. If turbulent diffusion in the intergalactic medium carries these fields over the separation of void galaxies, they could explain the void fields without requiring primordial fields and, consequently, physics beyond the SM. If primordial magnetic fields are present, these dipole fields would modify their signature significantly, requiring a careful subtraction of this astrophysical background.

In setup $A$ of \cref{tab:input}, with one percent of the magnetic energy in the dipole component, the characteristic field reaches about \(\SI{3e-13}{G}\) at the scale of \(1\unit{Mpc}\), and much smaller dipole fractions would still suffice to explain the blazar observations. The dependence on the galactic parameters is given by \cref{eq_Bcenter} and shown by the purple lines in \cref{fig:Bfield}. The white-noise spectrum provides a distinctive signature of the dipole contribution, which multiple halo measurements such as that of \Cite{tex:webar25} could test.

The present results can be extended in several directions. An independent determination of the effective turbulent diffusivity in void environments would establish whether the propagation distance is indeed sufficient. A realistic distribution of galaxies, including those within the voids, together with a distribution of dipole moments across galaxy types informed by cosmological simulations with astrophysical magnetogenesis~\cite{ref:Martin-Alvarez22}, would refine the amplitude and spatial structure of the resulting fields, while intrinsic alignments of galaxies and correlations among matter overdensities would modify the spectral shape. The relative motion of galaxies with respect to the surrounding gas, which could distort their external fields into extended wakes, is also not included here. Finally, quadrupolar components of galactic fields, whose toroidal part decays more slowly than the dipole in a conducting medium~\cite{Ghosh:2025bqp,Brandenburg:2025utr}, could provide an additional contribution.

\section*{Acknowledgements}

We thank Sergio Martin Alvarez, Rainer Beck, Abhijit Bendre, Axel Brandenburg, Chiara Caprini, Glennys Farrar, Andrii Neronov, Pranjal Ralegankar, Dmitri Semikoz, G\"unter Sigl and Kandu Subramanian for important comments on the first draft. Computations were performed on the Marvin cluster at University of Bonn.

\appendix

\section{Impact of model parameters}
\label{sec:sigmaB}

\begin{figure}
    \centering
    \includegraphics[width=\linewidth]{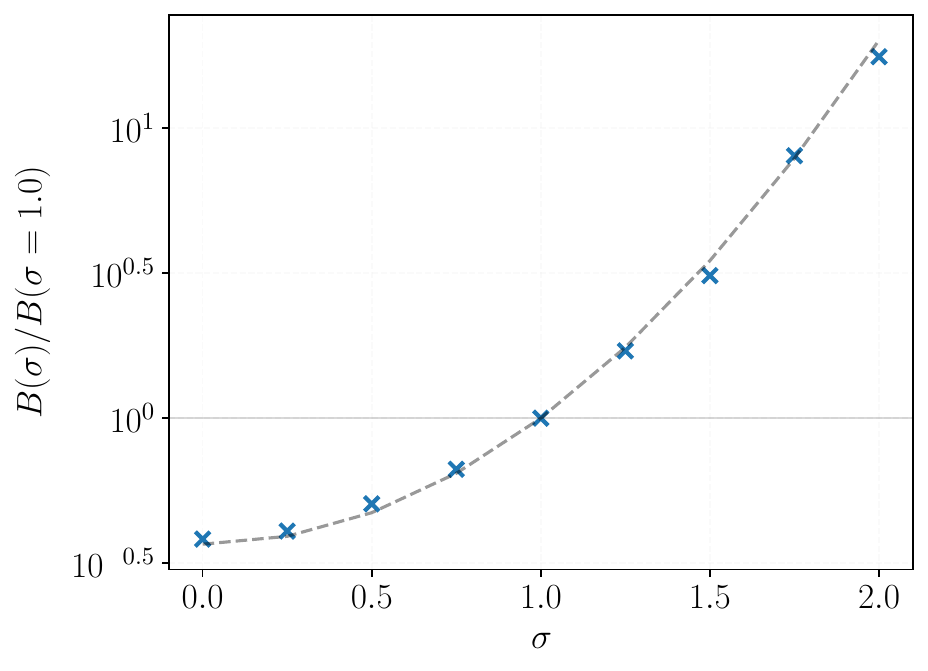}
    \caption{Dipole magnetic field in the void as a function of the standard deviation \(\sigma\) of the distribution of the galactic magnetic field, \(\ln \left(B\sub{g}/3\unit{\mu G}\right) \sim \mc{N} \left(0, \sigma^2\right)\), with the other parameters given by setup $A$ in \cref{tab:input}. The results are normalized to the case of \(\sigma = 1\), and the geometric mean of the corresponding ratio is taken across wavelengths. The dashed grey line shows the analytic expectation \(\log_{10}[B(\sigma)/B(1)] = (\sigma^2-1)/\ln 10\).}
    \label{fig:sigmaB}
\end{figure}

The expected strength of the dipole field in a void depends on the distribution assumed for the galactic fields in \cref{eq:bg}. Because the spectrum is a sum over independent dipoles, \(P\sub{B} \propto \sum_i B\sub{g,i}^2\), and for the lognormal distribution of \cref{eq:bg}, \(\langle B\sub{g}^2\rangle^{1/2} = \mu\sub{B}\,e^{\sigma^2}\). Relative to \(\sigma = 1\), the void field therefore scales as \(\log_{10}[B(\sigma)/B(1)] = (\sigma^2-1)/\ln 10\). Varying the standard deviation \(\sigma\) while keeping the other parameters as in setup $A$ of \cref{tab:input}, we find that the numerical results agree with this expression with an RMS deviation of about 8\% over the range shown in \cref{fig:sigmaB}, consistent with the statistical scatter expected from a finite number of galaxies and realizations.

\section{Convergence of the ensemble average}
\label{sec:convergence}

\begin{figure}
    \centering
    \includegraphics[width=\linewidth]{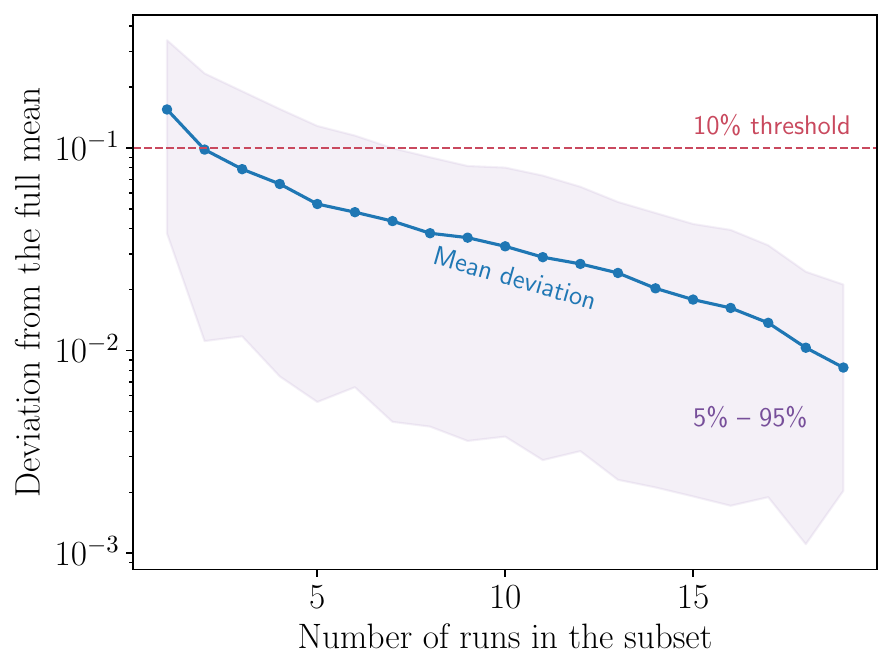}
    \caption{If we define the mean of twenty ensembles as the \textit{full mean}, and then consider various subsets, each containing $n$ ensembles, we can calculate the deviation of the mean of each subset from the \textit{full mean}. The spread of this deviation from the \textit{full mean} for the 95\% of the ensembles is shown as the purple region for different values of \(n\), with the horizontal axis representing \(n\). The mean deviation of all the subsets with a given $n$ is shown as the blue line. The mean deviation from the \textit{full mean} falls below 10\% as soon as we consider at least two ensembles, whereas the 95th-percentile deviation falls below 10\% for seven ensembles in the subset.
    }
    \label{fig:convergence}
\end{figure}

To calculate the ensemble average, different realizations of a void (setup $A$ in \cref{tab:input}) are considered, and the root mean square of the magnetic field is calculated from those realizations. Each realization of a void consists of a specific realization of 204 surrounding galactic dipoles modeled as stochastic variables. Each of these dipoles has a direction that is uniformly distributed over the surface of a sphere and a magnitude given by \(B\sub{dp} = \sqrt{f\sub{dp}} B\sub{g} = 0.1 B\sub{g}\), where \(B\sub{g}\) is assumed to have the distribution given in \cref{eq:bg}. We find that the ensemble average converges to a stable value with less than 10\% scatter after only two ensembles, which demonstrates the robustness of the averaging over 204 galaxies. To test this, we form subsets of \(n\) ensembles, with \(n\) ranging from two to nineteen, and compute the mean of each subset, using up to 500 distinct subsets for each \(n\), or all \(\binom{20}{n}\) subsets when fewer exist. For each subset mean, we then compute its RMS deviation across wavenumbers from the \textit{full mean} of the twenty ensembles. The spread of this deviation for 95\% of the ensembles is shown as the purple region in \cref{fig:convergence}. The mean deviation for a subset of n ensembles is shown as the blue line. The vertical axis is normalized to the \textit{full mean} of the twenty ensembles so the tick-labels represent relative deviation from the \textit{full mean}.

Expectedly, the deviation drops with increasing number of ensembles in the subset, as shown in \cref{fig:convergence}. If we just consider two ensembles, the mean deviation from the \textit{full mean} of twenty ensembles is already below 10\%, suggesting that twenty ensembles yield very close to the true mean of the system.
%

%

\end{document}